\documentclass[useAMS,fleqn, usenatbib, final]{mnras}

\usepackage{amssymb}
\usepackage{amsmath}
\usepackage{array}
\usepackage{graphicx}
\usepackage{color,units}
\usepackage[dvipsnames]{xcolor} 
\usepackage{aas_macros}
\usepackage{xspace}
\usepackage{dcolumn}
\usepackage{longtable}
\usepackage[normalem]{ulem} 
\usepackage{subfigure}
\usepackage[T1]{fontenc}
\usepackage{hyperref}
\graphicspath{ {./} }

\newcommand{\num}{1}
\newcommand{\portsmouth}{University of Portsmouth, Portsmouth, PO1 3FX, United Kingdom}
\newcommand{\bilby}{{\sc{bilby}\,}}
\newcommand{\pbilby}{{\sc{parallel-bilby}\,}}
\newcommand{\lalinference}{{\sc{LALInference}\,}}

\title[Inferring the properties of LISA sources with \bilby]{\bilby in space: Bayesian inference for transient gravitational-wave signals observed with LISA}

\author[Hoy and Nuttall]{
\parbox{\textwidth}{
C.~Hoy$^{\num}$\thanks{charlie.hoy@port.ac.uk},
L.~K.~Nuttall$^{\num}$
}
\vspace{0.2cm}\\
$^1$\portsmouth\\
}
\date{Accepted XXX. Received YYY; in original form ZZZ}
\pubyear{2023}

\begin{document}
\label{firstpage}
\pagerange{\pageref{firstpage}--\pageref{lastpage}}
\maketitle

\begin{abstract}
The Laser Interferometer Space Antenna (LISA) is scheduled to launch in the mid 2030s, and is expected to observe gravitational-wave candidates from massive black-hole binary mergers, extreme mass-ratio inspirals, and more. Accurately inferring the source properties from the observed gravitational-wave signals is crucial to maximise the scientific return of the LISA mission. \bilby, the user-friendly Bayesian inference library, is regularly used for performing gravitational-wave inference on data from existing ground-based gravitational-wave detectors. Given that Bayesian inference with LISA includes additional subtitles and complexities beyond it's ground-based counterpart, in this work we modify \bilby to perform parameter estimation with LISA. We show that full nested sampling can be performed to accurately infer the properties of LISA sources from transient gravitational-wave signals in a) zero-noise and b) idealized instrumental noise. By focusing on massive black-hole binary mergers, we demonstrate that higher order multipole waveform models can be used to analyse a year's worth of simulated LISA data, and discuss the computational cost and performance of full nested sampling compared with techniques for optimising likelihood calculations, such as the heterodyned likelihood.
\end{abstract}

\section{Introduction}

Gravitational-wave (GW) signals are regularly observed with existing ground-based GW detectors~\citep{LIGOScientific:2014pky,acernese2014advanced,KAGRA:2020tym}. By focusing on the Hertz to kiloHertz range of the GW spectrum, ground-based GW detectors have observed $\sim 100$ signals from neutron star and stellar-mass black hole coalescences~\citep{LIGOScientific:2021djp,Nitz:2021zwj,venumadhav2020new,zackay2019highly,zackay2019detecting,Mehta:2023zlk}, with many more expected during the ongoing fourth GW observing run. To probe lower frequencies, GW detectors must transition to space, as unavoidable terrestrial noise means that ground-based GW detectors will never be sensitive to the milliHertz frequency range~\citep{Danzmann:1996da}.

The Laser Interferometer Space Antenna (LISA) is designed to explore the milliHertz region of the GW spectrum, and is scheduled to launch in the mid 2030s~\citep{LISA:2017pwj}. This frequency range is rich in sources, with massive black hole binaries (MBHBs)~\citep{Klein:2015hvg}, extreme mass ratio inspirals~\citep{Glampedakis:2002cb,Babak:2006uv,Babak:2017tow} and galactic white dwarf binaries~\citep{Nelemans:2001hp} all expected. Inferring the properties of these systems from the observed gravitational-wave signals will e.g. provide stringent tests of general relativity in the strong-field regime, help constrain the Hubble's constant, and investigate the effect of dark matter on GW propagation~\citep{Amaro-Seoane:2012aqc}. Unfortunately, inferring the properties of LISA sources includes additional subtitles and complexities beyond it's ground-based counterpart. For instance, unlike existing ground-based GW detectors that can be assumed static throughout the observed GW signal, the LISA observatory is continuously orbiting the Sun. This affects e.g. waveform model generation as well as the commonly used assumption that the noise is stationary and uncorrelated. Accurately inferring the properties of LISA sources is crucial to maximize the scientific output of the LISA mission. 

\bilby, the user-friendly Bayesian inference library, is designed to infer the source properties from the observed GW signal~\citep{Ashton:2018jfp,Romero-Shaw:2020owr}. \bilby has been extensively used to analyse data from ground-based GW detectors, including individual GW observations~\citep[e.g.][]{LIGOScientific:2021djp}, population properties using hierarchical Bayesian inference~\citep[e.g.][]{KAGRA:2021duu,Saleem:2021vph,2019PhRvD.100d3030T}, tests of general relativity~\citep[e.g.][]{LIGOScientific:2021sio} and more~\citep[e.g.][]{You:2020wju,Powell:2022nrs,Hubner:2019sly,Dietrich:2020efo,Hoy:2022tst}. Given that \bilby is regularly used with a proven track record for producing reliable results, it is natural to extend \bilby to perform Bayesian inference on LISA sources. It also has the advantage of trivially interfacing with \pbilby~\citep{Smith:2019ucc}, which reduces wall-time through massively parallel Bayesian inference, as well as numerous methods for reducing overall cost~\citep[e.g.][]{Cornish:2010kf,Zackay:2018qdy,Cornish:2021lje,Morisaki:2023kuq,Morisaki:2021ngj,Williams:2021qyt,Krishna:2023aaa}.

In this paper, we outline our developments to \bilby to perform Bayesian inference on (simulated) LISA data. We demonstrate that full Bayesian inference can be performed on the three time-delay-interferometry (TDI) LISA observables~\citep{Tinto:2002de} for a range of waveform models. By focusing on MBHB mergers, we show that \bilby accurately infers the source properties in zero-noise, as well as idealized instrumental noise. We compare the standard likelihood commonly used in GW astronomy against optimised likelihoods designed to reduce the computational cost~\citep{Cornish:2010kf,Zackay:2018qdy,Cornish:2021lje,Krishna:2023aaa}. We show that although the latter reduces the CPU time by a factor of $\sim 10^{2}$, a small bias is observed. We finally conclude by showing that \bilby can perform Bayesian inference on LISA data with higher order multipole waveform models, and discuss how the inclusion of higher order multipoles can help break degeneracies in the parameter space. Although the work presented here focuses on MBHBs, our implementation is generic and can in principle perform Bayesian inference for any transient GW signals in LISA data; our only requirement is GW models for more exotic systems.

We note that there already exists methods to perform Bayesian inference on LISA sources: {\sc{balrog}}, a software package under development for LISA data analysis~\citep[see e.g.][]{Klein:2022rbf,Pratten:2022kug}, {\sc{pycbc-inference}}~\citep{Weaving:2023fji}, numerous algorithms that employ simplified fisher-matrix approximations~\citep[see e.g.][]{Vecchio:2003tn,Berti:2004bd,Lang:2006bsg,Arun:2008zn} and more~\cite[see e.g.][]{Cornish:2021smq,katz2022fully,Littenberg:2023xpl}. However, our implementation is generic, includes additional waveform models defined in {\sc{LALSimulation}}~\citep{lalsuite}, has the ability to interface with a diverse range of samplers, and builds upon the modularity and ease of using \bilby, meaning that it facilitates use by beginners.

The paper is organised as follows: in Section~\ref{sec:Bayesian} we review Bayesian inference, describe the likelihoods typically used in gravitational-wave astronomy, and introduce the \bilby package. In Section~\ref{sec:bilby_changes} we outline our modifications to \bilby to enable Bayesian inference on LISA data. In Section~\ref{sec:results} we validate our changes and confirm the effectiveness of \bilby for analysing LISA data. We infer the properties of MBHBs in zero-noise in Section~\ref{sec:zero_noise}, and idealised instrumental noise in Section~\ref{sec:gaussian_noise}. Results of the analyses presented in this paper are provided as accompaniments to this paper.

\section{Bayesian inference} \label{sec:Bayesian}

Bayesian inference is the process of estimating the parameters, $\boldsymbol{\theta}$, of a model, $m$, given the observed data $d$. These properties are represented by the model-dependent posterior probability density function (PDF), which is calculated through Bayes' theorem,

\begin{equation} \label{eq:bayes}
    p(\boldsymbol{\theta} | d, m) = \frac{p(\boldsymbol{\theta} | m)\, p(d | \boldsymbol{\theta}, m)}{\mathcal{Z}}.
\end{equation}
Here, $p(\boldsymbol{\theta} | m)$ is the probability that the model has parameters $\boldsymbol{\theta}$ given the model $m$, otherwise known as prior, $p(d | \boldsymbol{\theta}, m)$ is the probability of observing the data $d$ given the model parameters $\boldsymbol{\theta}$ and model $m$, otherwise known as the likelihood, and $\mathcal{Z}$ is the probability of observing the data given the model $m$, otherwise known as the evidence.

Although an analytic form of the prior and likelihood is known, it is often not possible to trivially evaluate Eq.~\ref{eq:bayes}. Consequently, stochastic sampling techniques, such as Nested sampling~\citep{Skilling2004,Skilling:2006gxv} or Markov-Chain-Monte-Carlo (MCMC)~\citep{metropolis1949monte}, are often used to draw samples from the unknown posterior PDF~\citep[although see e.g.][for alternative methods]{Pankow:2015cra,Lange:2018pyp,Delaunoy:2020zcu,Green:2020hst,Chua:2019wwt,Green:2020dnx,Dax:2021tsq,Gabbard:2019rde,Tiwari:2023mzf,Fairhurst:2023idl}.

In GW astronomy, the noise in the GW detector is assumed stationary and Gaussian, and the Whittle likelihood~\citep{whittle1951hypothesis} is often used~\citep{Romano:2016dpx,Thrane:2018qnx}. Here a parameterised GW model, $m$, is evaluated for a set of parameters, $\boldsymbol{\theta}$, and compared with the GW strain data, $d$. For a network of GW detectors, a Whittle likelihood is assumed for each case, and the full likelihood is obtained by taking the product of the single-detector likelihoods~\citep[see e.g.][for details]{Veitch:2014wba,Ashton:2018jfp}. This inherently assumes that the noise is uncorrelated in each detector. Of course, the GW observed in different detectors will have small differences owing to their relative position on Earth or in space. An additional step is therefore required to calculate the expected GW as seen in each of the detectors in the network.

As the Whittle likelihood is typically computationally expensive to evaluate, the heterodyned likelihood was introduced~\citep{Cornish:2010kf,Zackay:2018qdy,Cornish:2021lje}. By assuming that the model evaluated at two high likelihood points is similar, and hence the ratio is a smoothly varying function, summary data for a fiducial point can be pre-computed, and likelihoods for surrounding points can be evaluated in a fraction of the time. It has been shown that the heterodyned likelihood can reduce the computational cost of GW Bayesian inference by a factor of $10^{4}$ compared to evaluating the Whittle likelihood~\citep{Zackay:2018qdy}. A downside of the heterodyned likelihood is that a fiducial point of high likelihood must be known \emph{a priori}.

\subsection{\bilby and modifications for LISA} \label{sec:bilby_changes}

Numerous packages have been developed to stochastically sample the unknown posterior PDF for GW signals~\citep[e.g.][]{Veitch:2014wba,Ashton:2018jfp,Biwer:2018osg}. Most notably, \lalinference is a Bayesian inference library that employs custom built Nested and MCMC algorithms, and was the main workhorse of the LIGO--Virgo--KAGRA (LVK) collaboration for many years~\citep{Veitch:2014wba}. \bilby was later introduced as an alternative tool, which prioritised modularity and ease of accessibility~\citep{Ashton:2018jfp}. Unlike \lalinference, \bilby interfaces with existing `off-the-shelf' samplers~\citep[e.g.][]{Foreman-Mackey:2012any,Handley:2015vkr,Speagle:2019ivv,Williams:2021qyt,Ashton:2021anp}. For the case of the nested sampler {\texttt{dynesty}}~\citep{Speagle:2019ivv}, \bilby incorporates additional modifications to enhance it's performance for GW problems. Since the third LVK GW observing run, \bilby has primarily been used to analyse data from ground-based GW detectors~\citep{LIGOScientific:2021sio,KAGRA:2021duu}.

Although \bilby has a proven track record of analysing data from ground-based GW detectors, there are several additional features that are required in order to infer the properties of LISA sources. These include implementing the LISA observatory into the {\texttt{bilby.gw.detector}\,} module, described in Section~\ref{sec:lisa}, and incorporating additional models for calculating GWs as seen by LISA, see Section~\ref{sec:models}.

\subsubsection{The LISA observatory} \label{sec:lisa}

LISA will consist of three spacecraft, each containing a free falling test mass. Lasers will be relayed between each of the spacecraft, and the distance between free falling test masses will be monitored over time. A major challenge is that the laser frequency noise will be several orders of magnitude larger than any change in the arm length due to expected GW signals~\citep{LISA:2017pwj}. For this reason, Time-delayed interferometery (TDI) was introduced as an additional post-processing step. TDI is a technique that combines data from three spacecraft into virtual interferometric observables, where the laser frequency noise is suppressed by several orders of magnitude, improving the sensitivity of LISA to GW signals. Numerous TDI observables have been proposed with different underlying assumptions. The second-generation TDI observables, $X$, $Y$ and $Z$, allowed for linearly changing arm lengths, making them the most applicable for realistic LISA scenarios~\citep{Tinto:1999yr,armstrong1999time,Estabrook:2000ef,Vallisneri:2005ji,Tinto:2020fcc,tinto2023second}. However, the second-generation TDI observables are correlated in their noise properties. For this reason, a linear transformation is applied to the $X$, $Y$ and $Z$ observables, forming a set of approximately uncorrelated TDI observables: $A$, $E$ and $T$~\citep{Prince:2002hp}.

We provide the {\texttt{LISA}} class to handle the LISA observatory. We treat LISA as a network of distinct virtual interferometers, each corresponding to separate TDI observables. Although our implementation is generic to any TDI observables, allowing for possible improvements beyond second-generation TDI, the likelihood for a network of virtual interferometers assumes the noise is uncorrelated in each detector. This paper therefore uses the $A$, $E$ and $T$ channels.

\subsubsection{Alternative GW models} \label{sec:models}

In order to evaluate the GW likelihood, a parameterised model must be evaluated for a given set of parameters, and compared with the GW strain data. Since it is sensible to perform Bayesian inference on the TDI observables, see Section~\ref{sec:lisa}, our parameterised GW model must calculate the expected GW in each of these channels.

An additional complication is that the GWs from expected LISA sources will likely remain observable for weeks, if not months. During this time, the LISA observatory will continuously change it's orientation with respect to the source, as a result of it's orbit around the Sun. This implies that when calculating the expected GW as seen by LISA, we must apply a time and frequency dependent transformation, $\mathcal{T}(f, t)$, to the source-frame GW~\citep{Marsat:2020rtl}. This is in contrast to ground-based GW detectors, which are assumed to be static during the whole GW signal. 

The {\texttt{lisa\_binary\_black\_hole}} function returns a dictionary of expected GWs observed in each TDI observable for a provided set of parameters $\boldsymbol{\theta}$. Since we argued that the $A$, $E$ and $T$ channels are the most logical for calculating the likelihood in Section~\ref{sec:lisa}, we currently only consider these TDI observables. Of course, this can be extended in future work. Our implementation calculates GWs for parameters provided in the solar system baricenter (SSB) or LISA frame. When the LISA frame is used, transformations are applied to map the parameters into the SSB frame. This is required in order to calculate $\mathcal{T}(f, t)$, which depends on the source location and polarization defined in the SSB frame~\citep{katz2022fully}. 

We interface with the {\texttt{BBHx}}~\citep{Katz:2020hku,Katz:2021uax,michael_katz_2021_5730688} and {\texttt{LALSimulation}}~\citep{lalsuite} libraries. {\texttt{BBHx}} is an open source software that generates the GW polarizations, calculates the time and frequency dependent transformations, $\mathcal{T}(f, t)$, and returns the expected GWs in the $A$, $E$ and $T$ channels. Currently {\texttt{BBHx}} implements the {\texttt{IMRPhenomD}}~\citep{2016PhRvD..93d4007K, Husa:2015iqa} and {\texttt{IMRPhenomHM}}~\citep{London:2017bcn} waveform models; both models assume black hole spins aligned with the orbital angular momentum, with the former considering only the leading order quadrupole contribution, while the latter allows for additional higher-order multipole content. The user can call the {\texttt{BBHx}} models through \bilby by passing a {\texttt{waveform\_approximant}} of {\texttt{BBHx\_IMRPhenomD}} or {\texttt{BBHx\_IMRPhenomHM}} respectively. In order to reduce computational cost, interpolation can be enabled in the {\texttt{BBHx}} waveform models by specifying the waveform argument {\texttt{direct=False}}. Here, a reduced set of frequencies are used to evaluate the waveform, controlled with the waveform argument {\texttt{length}}, and interpolation is used to provide the waveform at the desired frequency points~\citep{Katz:2020hku,Katz:2021uax,michael_katz_2021_5730688}; by default {\texttt{length = 1024}}. In this work, we do not use interpolation when using the {\texttt{BBHx}} waveform models, in order to reduce the possibility of interpolation errors. If the user would like to use additional models beyond those included in {\texttt{BBHx}}, we use the {\texttt{LALSimulation}} library to generate GW polarizations, and use {\texttt{BBHx}} functions to project the GW polarizations into the $A$, $E$ and $T$ channels. This allows us to interface with e.g. the cutting edge binary black hole model {\texttt{IMRPhenomXPHM}}~\citep{Pratten:2020ceb}.

\begin{figure}
    \includegraphics[width=0.48\textwidth]{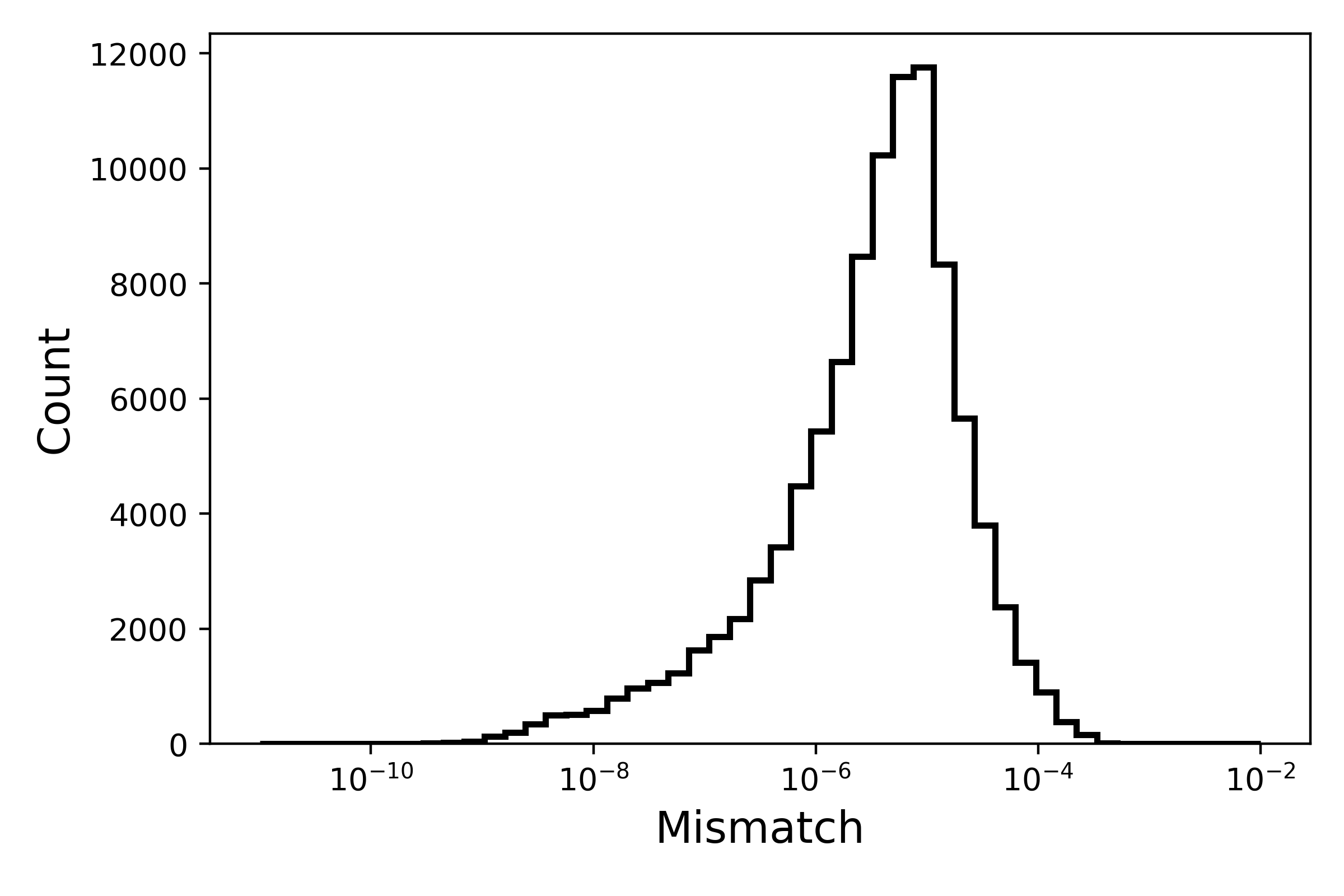}
    \caption{Mismatches between $10^{5}$ GW signals calculated using the {\texttt{BBHx}} and {\texttt{LALSimulation}} implementations of {\texttt{IMRPhenomD}}, when projected onto the A TDI channel.}
    \label{fig:mismatch}
\end{figure}

To validate our {\texttt{LALSimulation}} implementation, we randomly drew $10^{5}$ MBHBs with chirp masses $1 < \mathcal{M}\, [10^{6}\, M_{\odot}] < 5.0$, mass ratios $0.25 < q < 1.0$, spins $0 < \chi < 0.99$ and luminosity distances $1 < d_{L} [\text{Gpc}] < 100$ (all other parameters randomly chosen from an unconstrained boundary), and calculated the expected GW in the $A$ TDI channel when using the {\texttt{BBHx}} {\texttt{IMRPhenomD}} implementation, and the {\texttt{LALSimulation}} {\texttt{IMRPhenomD}} implementation. We quantify the difference between GWs by the mismatch, which is defined as,

\begin{equation}
    \mathfrak{M} = 1 - \max_{\phi, t}\left[4 \, \mathrm{Re} \int_{0}^{\infty} df 
    \frac{\tilde{a}(f) \tilde{b}^{\ast}(f)}{S(f)}\right]
\end{equation}
where $a$ and $b$ are the GWs calculated using {\texttt{BBHx}} and {\texttt{LALSimulation}} respectively, $S$ is the one-sided power spectral density (PSD), and the maximisation is over the phase $\phi$ and time $t$\footnote{By maximising over time, we are implicitly assuming that the LISA response function is not varying in time across the duration of this signal. However, because the majority of the signal-to-noise ratio from MBHBs occurs in the final day before merger~\citep{Marsat:2020rtl}, it is safe to assume that LISA is approximately stationary during this time period.}. The mismatch ranges between $0$ and $1$, where a mismatch of $0$ implies the two GWs are identical. As shown in Figure~\ref{fig:mismatch}, we find a median mismatch of $4.0\times 10^{-6}$, with a large tail extending to smaller mismatches. Interestingly, we find that in general larger mismatches are correlated with smaller chirp masses, where the signal durations are longer. Ideally all mismatches would be close to numerical precision, but we point out that the average mismatch remains two orders of magnitude lower than the mismatch between {\texttt{IMRPhenomD}} and numerical solutions to general relativity~\citep{2016PhRvD..93d4007K}. It is well known that the mismatch can be mapped to an approximate signal-to-noise ratio (SNR) at which two signals are indistinguishable at 90\% confidence~\citep{Baird:2012cu}. Assuming eleven degrees of freedom (for example the two masses and spins of both black holes, sky location, luminosity distance, inclination, phase, polarization and merger time), we find that on average, the two approaches are indistinguishable for SNRs less than $\sim 1400$, and $70\%$ of signals are indistinguishable for SNRs less than $\sim 1000$. Therefore although there may be convention differences between the two approaches, which become more evident at larger durations, it is unlikely to cause significant differences when performing Bayesian inference on most MBHB systems. We are therefore confident that our {\texttt{LALSimulation}} implementation is reasonable.

\section{Results} \label{sec:results}

To test our developments to \bilby to perform Bayesian inference on LISA data, we analyse MBHB signals. We choose signals that are similar to those expected when LISA is online. For cases where the true values are known, systematic biases can be studied.

We perform Bayesian inference with the {\texttt{dynesty}} Nested sampler~\citep{Speagle:2019ivv}, and compare the computational cost and performance of evaluating the full Whittle likelihood with the heterodyned likelihood. Since evaluating the full Whittle likelihood is likely computationally expensive, we perform Bayesian inference with \pbilby. We use standard \bilby when evaluating the heterodyned likelihood, and set the true values as the fiducial point. Of course, in reality the true values are unknown, meaning that this is an idealised case. For all analyses we use 1000 live points, the bilby-implemented {\texttt{rwalk}} sampling algorithm with an average of 60 accepted steps per MCMC, and evaluate the likelihood for frequencies $10^{-4}\, \mathrm{Hz} \leq f \leq 10^{-1}\, \mathrm{Hz}$. Sampling is performed in the LISA frame, as it is the most natural for LISA data analysis tasks, and we coherently analyse the $A$, $E$ and $T$ channels. All analyses use uninformative and wide priors, and we sample in the standard parameters: chirp mass, mass ratio, spin magnitudes, inclination angle, phase, ecliptic latitude, ecliptic longitude, polarization and merger time. We reduce the computational cost by analytically marginalizing over the luminosity distance of the source, and reconstructing the posterior in post-processing, see~\cite{Thrane:2018qnx} for details.

To compare posterior distributions obtained with the heterodyned and full Whittle likelihoods, we use the Jensen-Shannon divergence (JSD)~\citep{61115}. The JSD ranges between $0\,\mathrm{bits}$ and $1\,\mathrm{bit}$, where a JSD=$0\,\mathrm{bits}$ (JSD=$1\,\mathrm{bit}$) implies statistically identical (distinct) distributions. The JSD has been regularly used in GW astronomy for comparing posterior distributions~\citep[see e.g.][]{LIGOScientific:2018mvr,LIGOScientific:2020ibl}. A general rule of thumb is that a JSD $<0.05\,\mathrm{bits}$ implies that the distributions are in good agreement~\citep{LIGOScientific:2018mvr}.

We analyse a zero-noise injection in Section~\ref{sec:zero_noise}, and an injection in mock LISA data with idealised Gaussian instrumental noise in Section~\ref{sec:gaussian_noise}. We obtained mock LISA data through the Sangria dataset~\citep{le_jeune_2022_7132178} released as part of the LISA data challenge (LDC). The LDC contains multiple challenges that will eventually cover all of the complexities associated with LISA data analysis. For this paper, we only consider the Sangria dataset as a proof of principle, and leave further challenges to future work. The PSDs used in this work were generated from the Sangria dataset using {\texttt{gwpy}}~\citep{gwpy} and the Welch method. Unlike \cite{Weaving:2023fji}, we assume Galactic binaries form part of the noise, and therefore we include them in our PSD generation. By constructing PSDs in this way we are inherently removing the time-dependence, both from the foreground Galactic binaries and from LISA's orbit around the Sun. This is therefore an idealised case, but we note that it will unlikely cause a significant difference for MBHBs, where the SNRs are expected to be large, and rapidly accumulating within the final day before merger~\citep{Marsat:2020rtl}. This assumption will need to be relaxed for signals of longer duration. We note that the foreground Galactic binary problem may be navigated by using a Global Fit approach~\citep[see e.g.][]{Littenberg:2023xpl}, and the PSD may be able to include a time-frequency dependence by using the methods outlined in e.g. ~\cite{Cornish:2020odn}.

\begin{figure}
    \includegraphics[width=0.48\textwidth]{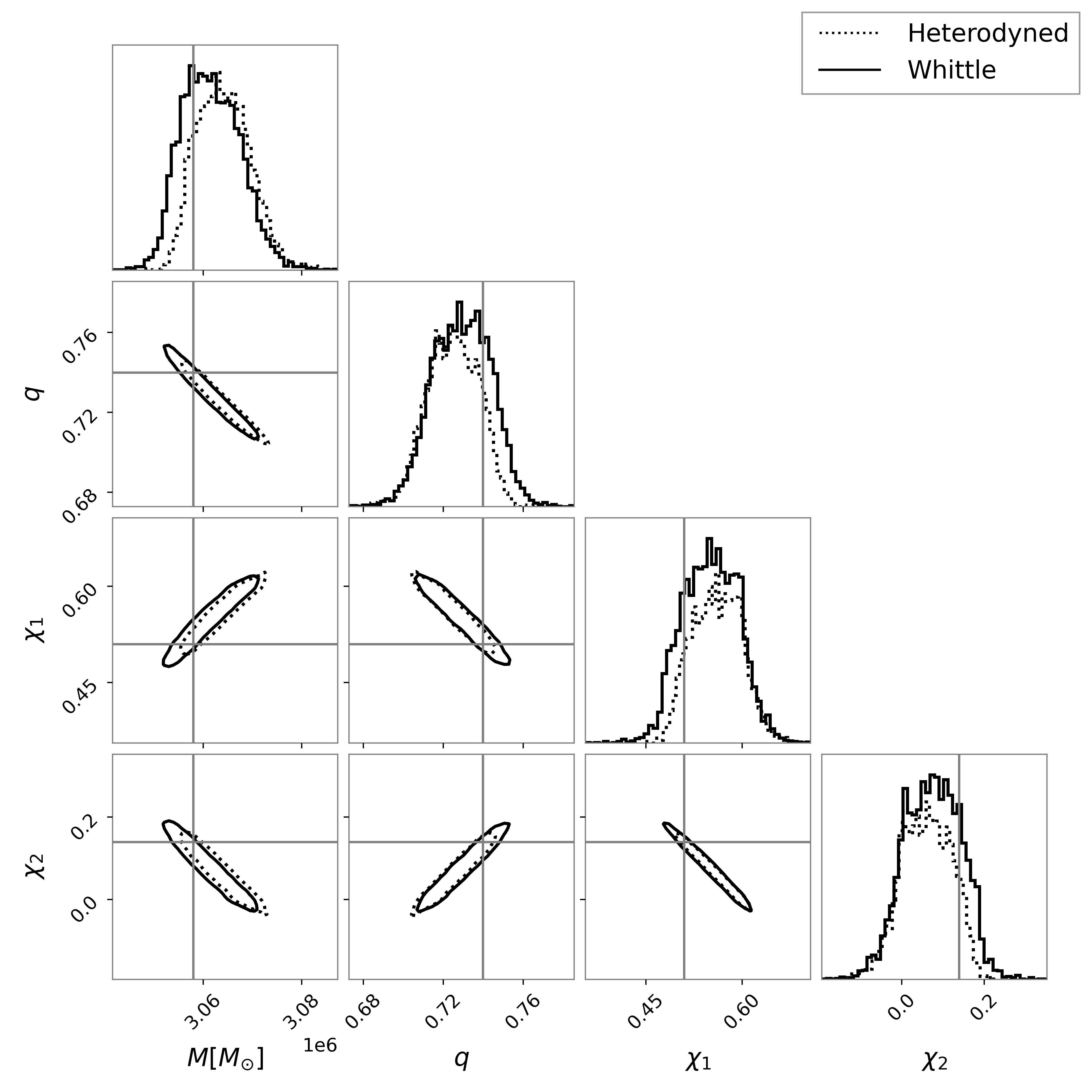}
    \caption{Inferred posterior distribution for the total mass $M$, mass ratio $q$, primary spin $\chi_{1}$ and secondary spin $\chi_{2}$ when analysing a zero noise injection with the {\texttt{BBHx\_IMRPhenomD}} model. Results were obtained when analysing the A, E and T TDI channels. The dotted and solid lines show the results when using the heterodyned and full Whittle likelihood respectively. The grey cross hairs show the true values, and the 2-dimensional contours show the 90\% confidence interval.}
    \label{fig:zero_noise}
\end{figure}

\subsection{Zero-noise Injections} \label{sec:zero_noise}

We first simulate a GW produced by a MBHB with total mass $M=m_{1} + m_{2}=3.1\times 10^{6}\, M_{\odot}$, mass ratio $q=m_{2}/m_{1}=0.74$, primary spin $\chi_{1}=0.51$ and secondary spin $\chi_{2}=0.14$. We assume the source has spins aligned with the orbital angular momentum and is located at an ecliptic latitude $\beta=2.8$ and ecliptic longitude $\lambda=2.3$, viewed at an inclination angle of $\iota=1.3\, \mathrm{rad}$ and luminosity distance $d_{L}=17\, \mathrm{Gpc}$. The phase and polarization were chosen to be $\phi=4.0$, $\psi=1.1$ respectively, and the merger time was set to be $t_{c}=6.7\times 10^{6}\, \mathrm{s}$. The injection is defined in the LISA frame, and has SNR $\sim 550$\footnote{The parameters were chosen to be similar to the zero-noise injection performed in~\cite{Weaving:2023fji}.}. We generate the simulated signal with {\texttt{BBHx\_IMRPhenomD}}, inject it into a month's worth of zero-noise data sampled at 5 second intervals, and perform Bayesian inference with the {\texttt{BBHx\_IMRPhenomD}} model for consistency.

\begin{figure*}
    \includegraphics[width=0.98\textwidth]{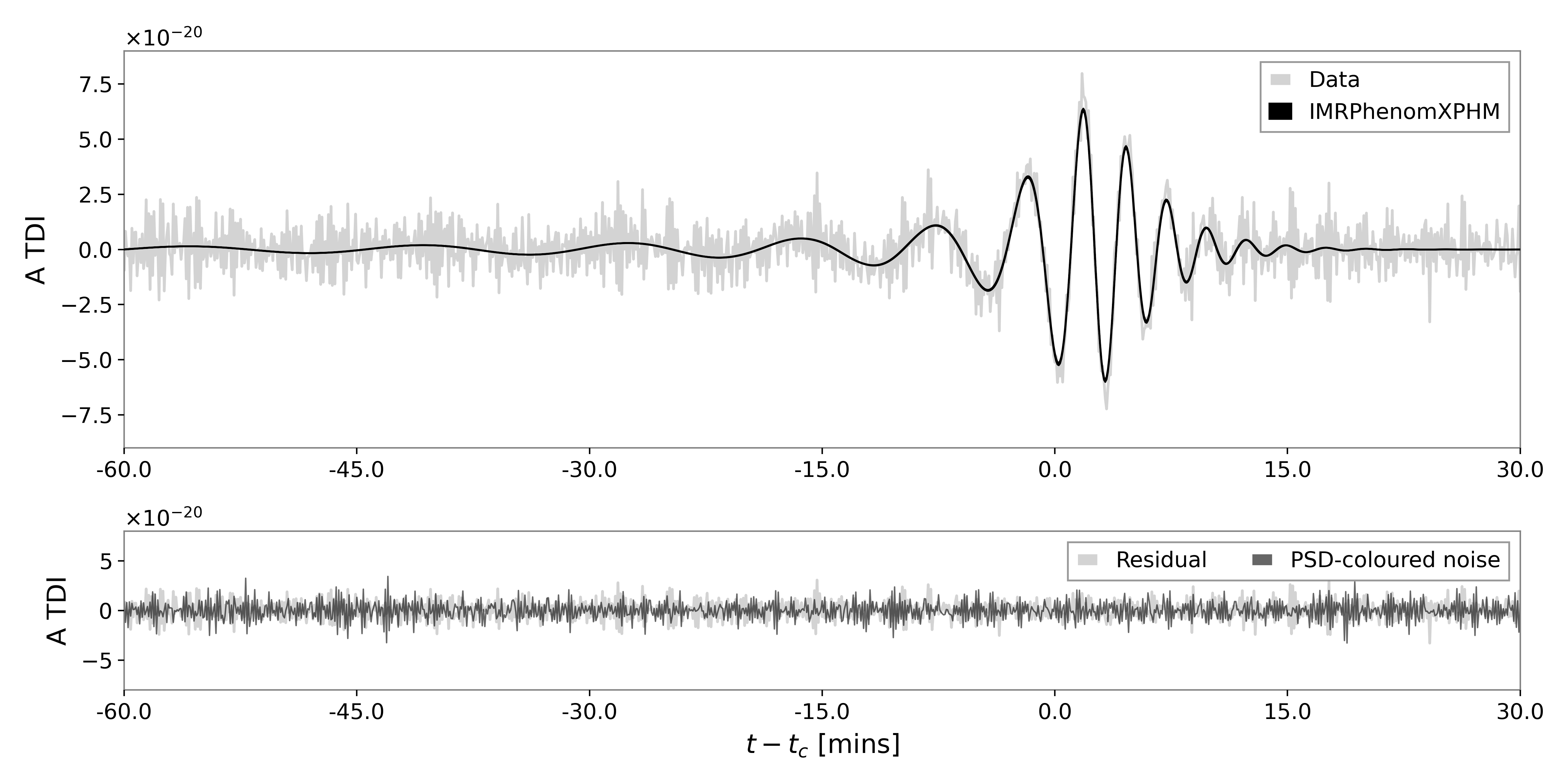}
    \caption{\emph{Top}: Comparison between the reconstructed GW signal (black) and the time-domain Sangria blind dataset (grey). The reconstructed GW signal was obtained by coherently analysing the A, E and T TDI channels using the aligned-spin limit of {\texttt{IMRPhenomXPHM}}, and the full Whittle likelihood. We shift the reconstructed GW signal and data by the inferred merger time as defined in the LISA frame $t_{c}$. The reconstructed GW signal is plotted as a band representing the 90\% credible region. \emph{Bottom}: Comparison between simulated gaussian noise which has been coloured by the power spectral density (black), and the residual after subtracting the reconstructed GW signal from the time-domain Sangria blind dataset (grey).  For simplicity, we only show the A TDI channel in both panels.
    }
    \label{fig:data}
\end{figure*}

In Figure~\ref{fig:zero_noise}, we compare the marginalized posterior distributions for the intrinsic parameters (total mass, mass ratio, primary spin and secondary spin) with the true values. In general we see that the full Whittle and heterodyned likelihoods recover the true values within the 90\% confidence interval. Although all Jensen-Shannon divergences are less 0.05 bits, we see that the Whittle likelihood results more accurately recover the true values, with a small shift relative to the heterodyned likelihood in all cases.

When considering extrinsic parameters (merger time, sky location, luminosity distance, inclination, phase, polarization), we expect to observe a bias in the sky location owing to well known degeneracies; for aligned-spin leading order quadrupole models, we expect to infer one of eight possibilities~\citep{Marsat:2020rtl}. Both likelihoods find support at the true ecliptic latitude, but only the heterodyned likelihood identifies the true ecliptic longitude mode, while the full Whittle likelihood favours the opposite sky mode. We see that both likelihoods recover a quad-modal polarization, with the full Whittle posterior shifted by $\pi/2$ compared to the heterodyned posterior -- this is consistent with favouring the opposite sky mode. Both likelihoods accurately recover the merger time and phase. Interestingly, only the Whittle likelihood observes a bias in the inclination angle and luminosity distance. Since we are using a model with spins aligned with the orbital angular momentum, and ignoring the effects of higher-order multipole content, we are unable to break the well known distance--inclination degeneracy~\citep[see e.g.][]{Marsat:2020rtl}. It is therefore expected to see a bias in both the distance and inclination angle, especially at such large SNRs. Given that the heterodyned likelihood uses the true values as the fiducial point, it is not surprising that it more accurately recovers the distance and inclination angle of the source.

We found that, although both analyses obtained $\sim 5,500$ posterior samples and evaluated the likelihood $\sim 200$ million times, the analysis that took advantage of the heterodyned likelihood was $\sim 10^{2}$ times faster and required $\sim 10$ times less memory than the full Whittle likelihood analysis; the full Whittle likelihood analysis completed in $O(10^{4})$ CPU hours and required $\sim 200\, \mathrm{GB}$ memory. We note, however, that the heterodyned likelihood requires prior knowledge of a high likelihood point, whilst the Whittle likelihood is completely general. To navigate this, previous works either used the true injected values, or they performed template-based matched-filter searches of the data prior to performing Bayesian inference, and they used the best fit template as the high likelihood point. We note that in some cases, the latter can lead to relatively poor fitting templates~\citep{Weaving:2023fji}. Standard implementations of the heterodyned likelihood may also not be appropriate for models that include the effects of higher-order multipoles~\citep{Leslie:2021ssu}. The full Whittle likelihood may therefore be more appropriate for analyses where a high likelihood fiducial point cannot be confidently found, or when models that include higher order effects are required. We leave an investigation that studies the impact of fiducial points, as well as potential algorithms for using the heterodyned posteriors as an initial guess for massively restricting the prior volume for full Whittle analyses, to future work.

\subsection{Sangria Blind Dataset} \label{sec:gaussian_noise}

We next analyse the Sangria Blind Dataset~\citep{le_jeune_2022_7132178} with {\texttt{IMRPhenomXPHM}}~\citep{Pratten:2020ceb}. {\texttt{IMRPhenomXPHM}} is the cutting edge binary black hole model, which includes spin-induced orbital precession~\citep{Apostolatos:1994mx} and higher order multipoles. In this analysis, we restrict black holes spins to be aligned with the orbital angular momentum, meaning that we ignore the effects of spin-induced orbital precession. Of course, we could have alternatively used {\texttt{IMRPhenomXHM}}~\citep{Garcia-Quiros:2020qpx}, which also includes higher order multipole contributions and similarly neglects the effects of spin-induced orbital precession. However, we chose to use the more general {\texttt{IMRPhenomXPHM}} model, as it is more likely to be used in future analyses; for instance investigating the impact of including mis-aligned black hole spins for breaking possible degeneracies. This dataset contains idealized Gaussian instrumental noise and simulated waveforms from Galactic white dwarf binaries and MBHBs with parameters derived from astrophysical models. The level of instrumental noise, the number of sources, and their parameters are not disclosed. Since the heterodyned likelihood requires a fiducial point, and the true injected parameters are unknown, we only use the full Whittle likelihood. We analyse a year's worth of data, sampled at 5 second intervals, and generate PSDs using all available data. The PSDs differ from those used in Section~\ref{sec:zero_noise}, but we similarly assume that Galactic binaries form part of the noise, and therefore we include them in our PSD generation.

As with the current \bilby implementations, we require an estimate for the merger time of the binary in order to define a suitable prior distribution for the merger time. This is typically provided from the matched-filter search pipelines. However, given that signal 0 can clearly be seen in the Sangria blind dataset, see Figure~\ref{fig:data}, it is trivial to estimate a suitable prior distribution that encompasses the merger. In future analyses of blind data, we would base our prior distributions on the best matching template from a LISA search pipeline, see e.g. \cite{Weaving:2023fji}.

We see remarkable agreement between the reconstructed GW obtained with {\texttt{IMRPhenomXPHM}} and the data, see Figure~\ref{fig:data}. When comparing the marginalized posterior distributions with those obtained by \cite{Weaving:2023fji} and {\texttt{BBHx\_IMRPhenomD}} (see Figs. 9 and 10 in \cite{Weaving:2023fji}), we see that our analysis a) more tightly constrains the intrinsic source properties, and b) obtains smoother (more gaussian-like) posterior distributions. This is expected as significant SNRs, as well as a lack of observed SNRs, in higher order multipoles can help break degeneracies~\citep[see e.g.][]{Marsat:2020rtl,Pratten:2022kug,Gong:2023ecg}. We can directly calculate the SNRs in higher order multipoles given the inferred posterior with \texttt{simple-pe}~\citep{Fairhurst:2023idl}. For this signal, we infer SNRs in the $(\ell, m) = (3, 3)$ and $(4, 4)$ multipoles to be $< 1$ ($\sim 0.6$ and $\sim 0.3$ respectively). The near-zero SNRs are consistent with the expectations from gaussian noise~\citep{Mills:2020thr}, and therefore we find no significant evidence for higher order multipole content in this signal. This is consistent with the fact that the GW signals in the Sangria Blind Dataset were constructed with {\texttt{IMRPhenomD}}~\citep{le_jeune_2022_7132178}.

When subtracting the reconstructed GW from the data, the residual signal is comparable with simulated gaussian noise coloured by the PSD, see Figure~\ref{fig:data}. This implies that the MBHB signal has been appropriately subtracted from the data, meaning that the residual can be used for further Bayesian inference analyses as part of a Global Fit style approach~\citep[see e.g.][]{Littenberg:2023xpl}. We note that do not expect perfect agreement between the residual and the simulated gaussian noise, since the simulated gaussian noise is dependent on the specific noise realisation chosen. 

The {\texttt{IMRPhenomXPHM}} analysis completed in $O(10^{5})$ CPU hours. Owing to the larger data duration considered, compared with the zero-noise injection presented in Section~\ref{sec:zero_noise}, more memory was required, $\sim 300\, \mathrm{GB}$. For comparison a typical analysis of 8 seconds worth of ground-based GW detector data sampled at $1/4096$ second intervals completes in $O(10^{3})$ CPU hours~\citep{Colleoni:2020tgc,Ramos-Buades:2023ehm}. The increase in computational cost is primarily driven by the waveform evaluation time and the long data durations considered. There are numerous methods available for reducing the computational cost without taking advantage of the heterodyned likelihood. For example, reduced order quadrature models~\citep{Canizares:2014fya,Qi:2020lfr,Morisaki:2023kuq}, adaptive frequency resolutions~\citep{Vinciguerra:2017ngf,Morisaki:2021ngj} and other techniques~\citep{Williams:2021qyt,Lee:2022jpn,Islam:2022afg,Wong:2023lgb,Pathak:2022ktt,Tiwari:2023mzf} have all been used to good effect in the past. Given that most of these methods are already implemented in the \bilby infrastructure, it is trivial to take advantage of these in future analyses. As discussed throughout this work, the SNR from MBHBs rapidly accumulates within the final day before merger~\citep{Marsat:2020rtl}. This implies that long data duration's may not be necessary into order accurately infer the properties of MBHBs. This will inherently reduce both the CPU and memory costs.

Although the scalability of \pbilby has been investigated in detail in \cite{Smith:2019ucc}, we provide a further analysis specific to the LISA context. We analysed the Sangria Blind Dataset with the aligned-spin limit of {\texttt{IMRPhenomXPHM}} using 128, 256, 512, and 768 CPUs. Owing to computational cost, we stopped the analyses after 12 hours wall time and compared the samplers progress. In general, we find that more CPUs increased a) the number of times the Whittle likelihood was called, b) the recorded iteration, and c) the number of proposals used. Specifically, when comparing the 768 and 128 CPU analyses, we find that the former called the likelihood $\sim 34$ times more than the latter, and it was on the $\sim 5000$th iteration compared to $\sim 3500$th. This suggests that running with more CPUs will reduce the overall wall time of the analysis.

\section{Discussion}

Inferring the properties of LISA sources is crucial for maximising the scientific output of the LISA mission. Although numerous algorithms exist for performing Bayesian inference, few are applicable for GW astronomy, and even fewer are capable of analysing LISA data. In this work, we introduce a modified version of \bilby, the user-friendly Bayesian inference library, to allow users to easily analyse LISA data with a suite of waveform models. We use \bilby as our starting point as it is already extensively used for analysing data from ground-based GW detectors, and has a proven track record for producing reliable results.

By focusing on MBHBs, we show that \bilby can perform Bayesian inference on LISA data and is capable of accurately inferring the source properties in zero-noise as well as idealized instrumental noise. We compare the performance of the standard likelihood used in GW astronomy, with an optimised likelihood designed to reduce the computational cost: the heterodyned likelihood. Although we find that analyses which use the full likelihood more accurately infer the source properties, analyses using the heterodyned likelihood are $\sim 10^{2}$ times faster. We note, however, that the heterodyned likelihood is dependent on knowing a point of high likelihood \emph{a priori}, and standard implementations may not be suitable for models with higher order multipole content. The full Whittle likelihood may therefore be preferred for many cases. We finally demonstrate that higher order multipole waveform models can be performed on LISA data. We highlight that the computational cost can likely be reduced by using techniques other than the heterodyned likelihood, already implemented into the \bilby infrastructure.

Our current implementation includes waveform models that are applicable for MBHBs. As we have maintained the modularity of \bilby, we anticipate future extensions of this work to include alternative waveform models~\citep{LISAConsortiumWaveformWorkingGroup:2023arg}.

\section*{Acknowledgements}
{
We would like to thank Jonathan Thompson for valuable comments on this manuscript, as well as Gregory Ashton, Ian Harry and Connor Weaving for continued discussions throughout this project. We are also grateful to Gregory Ashton and Colm Talbot for their advice regarding merging our modifications into the main \bilby code-base. We thank the UKRI Future Leaders Fellowship for support through the grant MR/T01881X/1. This work used the computational resources provided by the ICG, SEPNet and the University of Portsmouth, supported by STFC grant ST/N000064, and the DiRAC@Durham facility managed by the Institute for Computational Cosmology on behalf of the STFC DiRAC HPC Facility (www.dirac.ac.uk). The equipment for the latter was funded by BEIS capital funding via STFC capital grants ST/P002293/1, ST/R002371/1 and ST/S002502/1, Durham University and STFC operations grant ST/R000832/1. DiRAC is part of the National e-Infrastructure. This work used \bilby$=$ v2.1.0, {\sc{bilby\_pipe}}\,$=$ v1.0.11, \pbilby\,$=$ v2.0.2, and {\sc{dynesty}}\,$=$ v2.1.1. {\sc{PESummary}}~\citep{Hoy:2020vys} and {\sc{matplotlib}}~\citep{2007CSE.....9...90H} were used for plotting.
\\
\\
\emph{Data availability statement}: In this work we analysed simulated and publicly available data~\citep{le_jeune_2022_7132178}. The results presented in this paper are publicly available online~\citep{data_release}.
}

\bibliographystyle{mnras}
\bibliography{main}

\bsp	
\label{lastpage}
\end{document}